\documentclass[namedreferences]{SolarPhysics}
\usepackage[optionalrh]{spr-sola-addons} 

\usepackage{graphicx,times,url,color} 

\newcommand{\etal}{{\it et al.}}
\newcommand{\eg}{{\it e.g.}}
\newcommand{\degr}{\mbox{$^\circ$}}%

\newcommand{\an}{     {\it Astron. Nachr.}}
\newcommand{\apj}{    {\it Astrophys. J.}}
\newcommand{\apjl}{   {\it Astrophys. J. Lett.}}

\newcommand{\mnras}{  {\it Mon. Not. Roy. Astron. Soc.}}
\newcommand{\nat}{    {\it Nature}}

\newcommand{\pasj}{   {\it Pub. Astron. Soc. Japan}}
\newcommand{\solphys}{{\it Solar Phys.}}
\newcommand{\ssr}{    {\it Space Sci. Rev.}}

\begin{document}
\begin{article}
\begin{opening}

\title{Time--Distance Helioseismology Data Analysis Pipeline for 
       Helioseismic and Magnetic Imager onboard Solar Dynamics 
       Observatory (SDO/HMI) and Its Initial Results}
\author{J.~\surname{Zhao}$^{1}$ \sep
        S.~\surname{Couvidat}$^{1}$ \sep
        R.S.~\surname{Bogart}$^{1}$ \sep
        K.V.~\surname{Parchevsky}$^{1}$ \sep
        A.C.~\surname{Birch}$^{2}$ \sep
        T.L.~\surname{Duvall} Jr.$^{3}$ \sep
        J.G.~\surname{Beck}$^{1}$  \sep
        A.G.~\surname{Kosovichev}$^{1}$ \sep
        P.H.~\surname{Scherrer}$^{1}$ }
\runningauthor{J.~Zhao \etal}
\runningtitle{Time-Distance Pipeline for HMI}
\institute{$^{1}$ W.W.~Hansen Experimental Physics Laboratory, 
           Stanford University, Stanford, CA 94305-4085, USA
           email: \url{junwei@sun.stanford.edu} \\
           $^{2}$ NorthWest Research Associates, CoRA Division, 
           3380 Mitchell Lane, Boulder, CO 80301, USA \\
           $^{3}$ Laboratory for Astronomy and Solar Physics, NASA Goddard 
           Space Flight Center, Greenbelt, MD 20771, USA 
          }

\begin{abstract}
The {\it Helioseismic and Magnetic Imager} onboard the {\it Solar Dynamics 
Observatory} (SDO/HMI) provides continuous full-disk observations of solar 
oscillations. We develop a data-analysis pipeline based on the time-distance
helioseismology method to measure acoustic travel times using HMI Doppler-shift
observations, and infer solar interior properties by inverting these 
measurements. The pipeline is used for routine production of near-real-time 
full-disk maps of subsurface wave-speed perturbations and horizontal 
flow velocities for depths ranging from 0 to 20 Mm, every eight hours. 
In addition, Carrington synoptic maps for the subsurface properties are 
made from these full-disk maps. The pipeline can also be used for selected 
target areas and time periods. We explain details of the pipeline 
organization and procedures, including processing of the HMI Doppler 
observations, measurements of the travel times, inversions, and constructions 
of the full-disk and synoptic maps. Some initial results from the pipeline, 
including full-disk flow maps, sunspot 
subsurface flow fields, and the interior rotation and meridional 
flow speeds, are presented. 
\end{abstract}


\end{opening}

\section{Introduction}
The {\it Helioseismic and Magnetic Imager} onboard the {\it Solar Dynamics 
Observatory} (SDO/HMI: \opencite{sch11}) observes the solar full-disk 
intensity, Doppler velocity, and vector magnetic field of the photosphere with 
high spatial resolution and high temporal cadence. Similar to the {\it 
Michelson Doppler Imager} (MDI: \opencite{sch95}), an instrument onboard 
the {\it Solar and Heliospheric Observatory} (SOHO), 
the HMI Dopplergrams are primarily used for helioseismic analysis 
to investigate the interior structure and dynamics of the Sun. 
Helioseismology data analysis pipelines are planned for near real-time 
analyses of the observations in order to provide the analysis 
results to the helioseismology and solar physics communities. The
time--distance analysis pipeline is one of the pipelines for 
local helioseismology studies, and other pipelines include ring--diagram
analysis and farside active region imaging. The time--distance pipeline is 
designed for routine production of nearly full-disk subsurface wave-speed 
perturbations and 
horizontal flow fields every eight hours, as well as synoptic flow maps for 
every Carrington rotation. It can also be used to analyze specific target
areas and time periods.

Time--distance helioseismology was first introduced by Duvall \etal\ (1993, 
1996), and it has developed rapidly since then. Different inversion 
techniques were introduced and tested. The LSQR algorithm, introduced by 
\inlinecite{kos96} and used later by \inlinecite{zha01}, solves the 
inversion problem in a least-squares sense in the spatial domain by an 
iterative approach. The Multi-Channel Deconvolution (MCD) method, introduced 
by \inlinecite{jac99} and widely used in later studies (\eg\ \opencite{cou04}),
solves the least-squares problems in the Fourier domain. Later, 
\inlinecite{cou05} applied a horizontal regularization procedure for 
this inversion technique. More recently, an optimally localized averaging 
(OLA) inversion scheme was introduced to study the solar subsurface flow 
fields \cite{jac08}.

Different types of sensitivity kernels, which describe the relationship 
between the travel times and interior properties, were also introduced and 
used in the time-distance inversion problems. \inlinecite{kos96} first used 
ray-path approximation kernels, \inlinecite{jen00} introduced
Fresnel-zone kernels; and \inlinecite{bir00}, \inlinecite{bir04}, and 
\inlinecite{bir07} investigated Born-approximation kernels for both 
sound-speed structures and flow fields. \inlinecite{cou06} compared 
subsurface sound-speed perturbation structures inferred from these 
different types of kernels, and found that the inversion results obtained
with the different kernels were basically consistent.

Important results on the solar interior properties have been obtained from 
the time--distance studies as well as from other local helioseismology
techniques. On global scales, poleward meridional flows
were found below the photosphere \cite{gil97}, and solar-cycle dependent
meridional flow variations were also investigated and discussed \cite{cho01, 
bec02, zha04}. On local scales, subsurface sound-speed perturbations and 
flow fields were derived for supergranulation \cite{kos97, duv97, duv00, 
sek07, jac08} and for sunspots \cite{kos00, giz00, zha01, cou06, zha10}. 
Additionally, time-distance helioseismology was used to detect the 
emergence of active regions before their appearances in the photosphere 
\cite{kos00, jen01, zha08}, to image large active regions on the farside
of the Sun \cite{zha07b, ilo09}, and to measure sound-speed perturbations in 
the tachocline \cite{zha09}. These results are important for space-weather 
forecasting and understanding the mechanisms for the generation
of solar magnetism. The time--distance helioseismology pipeline analyses, 
based on the high spatial-resolution and high temporal-cadence observations 
from HMI, will greatly advance our knowledge of the interior processes 
and their connections with solar activity above the photosphere.

However, one should keep in mind that the physics of solar oscillations 
in the turbulent magnetized plasma is very complicated, and that the 
helioseismic techniques are still in the process of being developed. 
Because of limited knowledge of the wave physics and complexity of 
the MHD turbulence, there may be systematic uncertainties in the local 
helioseismology inferences, particularly in strong magnetic-field regions 
of sunspots. For example,
Lindsey and Braun (2005a, 2005b) argued that the outgoing and ingoing 
travel time asymmetries observed in sunspot areas might be caused 
by a ``shower-glass effect". \inlinecite{sch05} found that the inclined
magnetic field in sunspot penumbrae might cause variations of measured 
acoustic travel times. \inlinecite{zha06} found that this inclined magnetic 
field dependent effect does not exist in the measurements obtained 
from the MDI intensity observations. Later, \inlinecite{raj06} found that the 
non-uniform acoustic power distribution in the photosphere also contributed 
to measured travel-time shifts in active regions if a phase-speed filtering 
procedure was applied. This effect was then studied by \inlinecite{par08} and 
\inlinecite{han08} numerically, and also by \inlinecite{nig10} analytically. 
More recently, \inlinecite{giz09} derived a sunspot's subsurface flow fields 
after applying ridge filtering, and their inferred results did not agree 
with the previously inverted results with a use of phase-speed filtering. 
Using high spatial-resolution observations from {\it Hinode},
\inlinecite{zha10} showed that the principal results on sunspot 
structure did not depend on the use of phase-speed filtering. However, 
significant systematic uncertainties in sunspot seismology remain and need 
to be understood, and these are being actively studied by the use of numerical
simulations. For a recent review of the sunspot seismology and uncertainties
see \inlinecite{kos10}.

\begin{figure}
\centerline{\includegraphics[width=0.59\textwidth,angle=270]{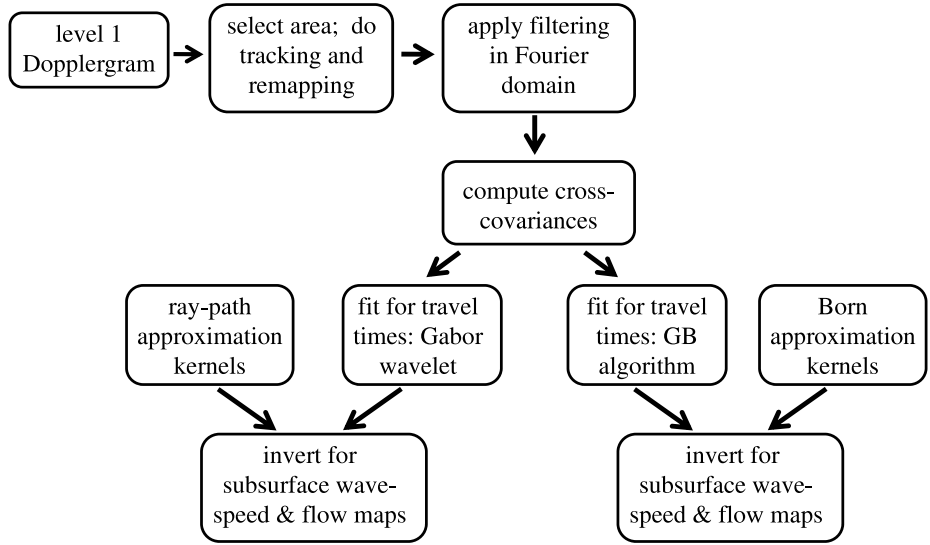} }
\caption{Flow chart for the HMI time-distance helioseismology data analysis 
pipeline.}
\label{flow_chart}
\end{figure}

Despite the ongoing discussions of accuracy of time--distance 
measurements and interpretation of inversion results, it is 
useful to provide the measured travel times and the inversion results 
for investigations of structures and flows below the 
visible surface of the Sun. As the flow chart of the pipeline in 
Figure~\ref{flow_chart} shows, we apply phase-speed filtering to the HMI data, 
compute cross-covariances of the oscillations, and use two different 
travel-time fitting procedures to derive the acoustic travel times. We 
then perform inversions using two different sets of travel-time sensitivity 
kernels, based on the ray-path and Born approximations, to infer subsurface 
wave-speed perturbations and flow fields, using the MCD inversion method. 
We provide online access to the measured travel times, full-disk subsurface 
wave-speed perturbations and flow maps calculated every eight hours. In 
addition, we provide synoptic flow maps for each Carrington rotation. 
In this article, we describe details of the acoustic travel time measurement 
procedure in Section~\ref{S2}, and the inversion procedure in Section~\ref{S3}. 
We describe the pipeline data products and present initial HMI results 
in Section~\ref{S4}, and summarize our work in Section~\ref{S5}.

\section{Acoustic Travel-Time Measurement}
 \label{S2}

\subsection{Tracking and Remapping}

The SDO/HMI continuously observes the full-disk Sun, providing Doppler 
velocity, continuum intensity, line-depth, line-width, and magnetic 
field maps with a 45-second cadence, and also vector magnetic field 
measurements with a cadence of 12 minutes. Each full-disk image has 
$4096\times4096$ pixels with a spatial resolution of 0.504 arcsec 
pixel$^{-1}$ ({\it i.e.}, approximately, 0.03 heliographic degree pixel$^{-1}$ 
at the solar disk center). The Doppler observations are primarily used 
for helioseismic studies. The observing sequences, algorithms for 
deriving Doppler velocity and magnetic field, and other instrument 
calibration issues are discussed by \inlinecite{sch11}.

As illustrated in Figure~\ref{flow_chart}, the primary input for the 
pipeline is Dopplergrams, although in principle, the HMI intensitygrams and
line-depth data can also be analyzed in the same manner. Users of the 
pipeline can select specific areas for analysis, preferably within 
$60\degr$ from the solar disk center. In practice, the users provide 
the Carrington longitude and latitude of the center of the selected 
area, and the mid-time of the selected time period, then the 
pipeline code selects an area of roughly $30\degr \times 30\degr$ 
centered at the given coordinate, and for a time interval of eight 
hours with the given time as the middle point. The data for 
this selected area and the time period are then tracked to remove
solar rotation, and remapped into the heliographic coordinates using 
Postel's projection (also known as azimuthal equidistant projection) 
relative to the given area center. Normally, 
the tracked area consists of $512 \times 512$ pixels with a spatial sampling of 
$0.06\degr$ pixel$^{-1}$; and the temporal sampling is the same as
the observational cadence. Cubic interpolation is used for the
pixels not located on the observational grid. 

\begin{figure}
\centerline{\includegraphics[width=0.75\textwidth]{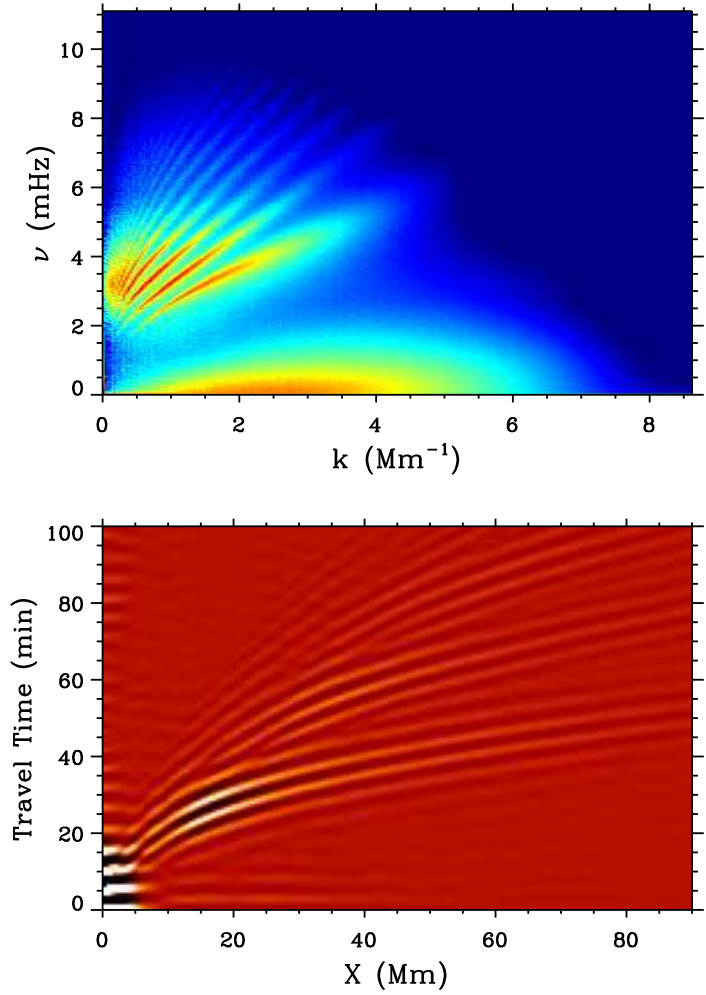} }
\caption{Typical power spectrum ($k-\omega$) diagram (upper) and 
time--distance (cross-covariance) diagram (lower) made from eight hours 
of HMI Doppler observations. }
\label{power_td}
\end{figure}

Figure~\ref{power_td} shows typical HMI $k$--$\omega$ and time--distance 
diagrams obtained from one tracked and remapped area. The selected area 
covers $30\degr$ in latitude and has an apparent differential 
rotation over this span. However, for fast computations, only one 
uniform tracking rate corresponding to the Snodgrass rotation rate 
at the center of this area is used.
The Snodgrass rotation rate is: $2.851 - 0.343 \sin^2\phi - 0.474 
\sin^4\phi$ $\mu$rad~s$^{-1}$, where $\phi$ is latitude \cite{sno90}. 
The uniform tracking rate of the selected area results in a differential 
rotation velocity in the inverted horizontal velocity fields. 
This differential rotation velocity is removed from the full-disk flow
maps after averaging over the whole Carrington rotation, 
and only the residual flow fields are given as the final results (see 
Section~\ref{S41}). But for the user-selected areas, the differential 
rotation is kept in the inversion results.

\begin{table}
\caption{Phase-speed filtering parameters used for the selected travel
distances (annulus ranges).}
\label{param_list}
\begin{tabular}{cccc}
\hline
annulus No. & annulus range & phase speed   & FWHM \\
            & (heliographic degree) & ($\mu$Hz/$\ell$) & ($\mu$Hz/$\ell$) \\
\hline
1 & 0.54 -- 0.78 & 3.40 & 1.0 \\
2 & 0.78 -- 1.02 & 4.00 & 1.0 \\
3 & 1.08 -- 1.32 & 4.90 & 1.25 \\
4 & 1.44 -- 1.80 & 6.592 & 2.149 \\
5 & 1.92 -- 2.40 & 8.342 & 1.351 \\
6 & 2.40 -- 2.88 & 9.288 & 1.183 \\
7 & 3.12 -- 3.84 & 10.822 & 1.895 \\
8 & 4.08 -- 4.80 & 12.792 & 2.046 \\
9 & 5.04 -- 6.00 & 14.852 & 2.075 \\
10 & 6.24 -- 7.68 & 17.002 & 2.223 \\
11 & 7.68 -- 9.12 & 19.133 & 2.039 \\
\hline
\end{tabular}
\end{table}

\subsection{Computing Cross-Correlations and Fitting for Travel Times}

Each tracked and remapped Dopplergram datacube is filtered in the 
3D Fourier domain. Solar convection and $f$-mode oscillation signals are 
removed first, and then phase-speed filtering is applied following the
procedures prescribed by \inlinecite{cou05}. For the travel-time 
measurements, for each central point we select 11 annuli with various 
radii and widths chosen from our past experience with MDI analyses. 
All of the phase-speed filtering parameters, including the central 
phase-speed, filter width, and the corresponding inner and outer 
annulus radii are presented in Table~1. The phase-speed filter is a 
Gaussian function of the wave's horizontal phase speed. It selects wave
packets traveling between the central points and the annuli for the 
selected distances. This filter helps improve the signal-to-noise ratio,
and also removes leakage from low-degree oscillations. After the 
filtering, the data are transformed back to the 
space--time domain for cross-covariance computations. Two different 
fitting methods are used to derive the acoustic travel times from 
the cross-covariances: a Gabor wavelet fitting \cite{kos97}, 
and a cross-correlation method based on seismology algorithms 
\cite{zha98} adopted 
by Gizon and Birch (2002; GB algorithm hereafter). A detailed description 
of the filtering procedure, the cross-covariance computations, the two types of 
travel-time fittings, comparison of the travel times derived from 
the two fitting methods, and the measurement error estimates 
is given by \inlinecite{cou11}. In particular, 
it has been shown there that the two definitions of the acoustic 
travel times and the two fitting methods give generally consistent results 
in the quiet-Sun regions, but may give different results in active 
regions. The differences are currently not well understood, but in 
the pipeline we implement both travel-time definitions.

In the Postel-projected maps, the exact distance between two arbitrary 
points cannot be calculated using the Cartesian coordinates. Thus, when an
annulus is selected around a given location, some additional computations are
needed to determine the exact great-circle distance between the points.
The formula to determine the great-circle distance is: 
\begin{equation} 
\Delta = \arccos ( \sin \theta_1 \sin \theta_2 + \cos \theta_1 \cos \theta_2
\cos (\phi_1 - \phi_2) ),
\end{equation} 
where $(\theta_1, \phi_1)$ and $(\theta_2, \phi_2)$ are the heliospheric
longitude and latitude coordinates for the two separate locations. 

To facilitate the inversions for subsurface flow fields, each annulus
is divided into four quadrants representing 
the North, South, East, and West directions \cite{kos97}.
So, for each annulus and each travel-time fitting method, we obtain 
six measurements of acoustic travel times, corresponding to the outgoing and 
ingoing, East-, West-, South-, and North-going directions. 
We then combine these travel times to obtain one map for the mean travel time 
and three maps for the travel-time differences. 
These travel times are: $\tau_\mathrm{mean}$ (average of 
outgoing and ingoing travel times), $\tau_\mathrm{oi}$ (difference of 
outgoing and ingoing travel times), $\tau_\mathrm{we}$ (difference of 
West- and East-going travel times), and $\tau_\mathrm{ns}$ (difference 
of North- and South-going travel times). These four travel-time 
maps for each annulus are archived and available through
the HMI Data Record Management System (DRMS).

\section{Subsurface Wave-Speed Perturbation and Flow Field Inversions}
 \label{S3}
The acoustic travel times are derived by two different fitting methods: 
the Gabor wavelet function and the GB algorithm. Then, as illustrated in 
Figure~\ref{flow_chart}, to infer the subsurface wave-speed perturbations and 
flow velocities, the Gabor-wavelet travel times are inverted 
using the ray-path approximation sensitivity kernels, and the GB travel 
times are inverted using the Born-approximation sensitivity kernels.
The Born-approximation kernels are calculated based with the filter 
and window parameters of the GB fits.

\subsection{Inversions}
 \label{S31}
Both the ray-path and Born-approximation kernels 
have been used in previous time--distance studies (\eg\ \opencite{zha01};
\opencite{cou06}). Details of the kernel calculations and their comparisons
will be given in a separate paper. 

We employ the MCD inversion method \cite{jac99}
with a horizontal regularization \cite{cou05}. For the wave-speed 
perturbation inversions, the linearized equation relating the measured 
mean travel times and the subsurface wave-speed perturbations are: 
\begin{equation} 
\delta \tau_\mathrm{mean}^{\lambda\mu\nu} = \sum_{ijk} A_{ijk}^{\lambda
\mu\nu} \delta s_{ijk},
\label{eq2}
\end{equation}
where $\delta s_{ijk}$ is the relative wave-speed perturbation 
$\delta c_{ijk}/c_{ijk}$ approximated by piece-wise constant functions 
on the inversion grid, and $A_{ijk}^{\lambda\mu\nu}$ is a matrix of 
the discretized sensitivity kernel. Here, $\lambda$ and $\mu$ label 
the coordinates of the central points of the annuli in the observed areas, 
$\nu$ labels the different annuli, and $i,j$, and $k$ are the indices 
for the discretized three-dimensional space for inversions. Usually, 
the horizontal coordinates of the inversion grid ($i$ and $j$ indices)
are selected at the same locations as the central travel-time measurement
points ($\lambda$ and $\mu$ indices). In the first-order approximation, the 
sensitivity kernels are calculated for a spherically symmetric solar 
model and do not depend on the position on the solar surface. Therefore, 
in this case Equation (\ref{eq2}) is actually equivalent to a convolution 
in the horizontal domain, which can be simplified as a direct multiplication 
in the Fourier domain: 
\begin{equation}
\delta\tilde{\tau}^{\nu} ( \kappa_{\lambda}, \kappa_{\mu} ) = \sum_{k}
\tilde{A}_{k}^{\nu} (\kappa_{\lambda}, \kappa_{\mu}) \delta \tilde{s}_{k}
(\kappa_{\lambda}, \kappa_{\mu}) ,
\end{equation}
where $\delta\tilde{\tau}$, $\tilde{A}$, and $\delta\tilde{s}$ are the 
2D Fourier transforms of $\delta\tau$, $A$, and $\delta s$, respectively; 
$k$ is the same as in Equation (2); $\kappa_{\lambda}$ and $\kappa_{\mu}$ 
are the wavenumbers in the Fourier domain corresponding to $\lambda$ 
and $\mu$ of the spatial domain. For each $(\kappa_\lambda, \kappa_\mu)$, 
the equation in the Fourier domain is a matrix multiplication:
\begin{equation} \label{eq}
d = G m,
\end{equation}
where 
$$d = \big\{ \delta \tilde{\tau}^{\nu} (\kappa_{\lambda}, \kappa_{\mu}) 
\big\},
\quad G= \big\{\tilde{A}_{k}^{\nu}(\kappa_\lambda, \kappa_\mu) \big\}, 
\quad m = \big\{\delta\tilde{s}_k (\kappa_\lambda, \kappa_\mu) \big\} .$$
Thus, we have a large number of linear equations describing the depth
dependence of the Fourier components, and each linear equation can be 
solved in the least squares sense. After all these equations are 
solved, and $m$ is obtained for each $(\kappa_\lambda, \kappa_\mu)$, 
the values of $\delta s_{ijk}$ are calculated by the inverse 2D 
Fourier transform. 

Equation (\ref{eq}) is ill-posed, and regularization is required to
obtain a smooth solution. The regularized least-squares algorithm
is formulated as:
\begin{equation}
\min \big\{ \Vert (d - G m) \Vert_2^2 + \lambda^2 (\kappa) \Vert L m \Vert_2^2
\big\} ,
\end{equation}
where $\Vert ... \Vert_2 $ denotes the L2-norm, $L$ is a regularization 
operator, and $\lambda (\kappa)$ is a regularization parameter. We choose
$L$ to be a diagonal matrix whose elements are the inverse of the square 
root of the spatial sampling $\Delta z$ at each depth. Such weighting is 
necessary because the grid in the vertical direction is chosen to be 
approximately uniform in the acoustic depth, meaning that the spatial sampling 
of deep layers is larger than the sampling of shallower layers. The regularization
parameter is taken in the form of $\lambda^2 ( \kappa ) = \lambda_v^2 + 
\lambda_h^2 (\kappa)$, where $\lambda_v$ and $\lambda_h$ are vertical 
and horizontal regularization parameters. The purpose of 
introducing $\lambda_h$ is to damp the high wavenumber components that 
may lead to noise amplification. Following the discussion of
\inlinecite{cou05}, we choose $\lambda_h = \lambda_2 \kappa^2$ with 
$\lambda_2$ as a constant. 

Because the regularization is applied in the Fourier domain, it is
quite difficult to use different regularization parameters for different
horizontal locations of the same region. Sometimes, different regularization
parameters are needed because in active regions the noise level may be 
different from the quiet-Sun regions, as we discuss in Section~3.3. Thus, 
it is necessary to implement into the analysis pipeline another inversion 
technique, the LSQR algorithm, which solves Equation~(2) in the space domain 
by an iterative approach. This implementation is currently under development. 

\subsection{Inversion Depth and Validation of Inversions}
 \label{S32}
For both the wave-speed 
and flow-field inversions, and for both the ray-path and Born-approximation 
inversions, we select a total of 11 inversion depths 
as follows: 0 -- 1, 1 -- 3, 3 -- 5, 5 -- 7, 7 -- 10, 10 -- 13, 
13 -- 17, 17 -- 21, 21 -- 26, 26 -- 30, and 30 -- 35 Mm. There is 
a total of 11 depth intervals. The inversion results provide 
the wave-speed perturbations and flow velocities averaged in these layers.
Due to the lack of acoustic wave coverage in the deep interior, 
the reliability of inversion results decreases with the depth. Thus, 
only inversion results for the depths shallower than 20 Mm are 
included in the pipeline output. This may change in the future when 
more confidence is gained in the deeper interior inversion results.

In recent years, several studies have been carried out to validate the 
time--distance measurements and inversions. To validate the derived subsurface 
flow fields, \inlinecite{geo07} and \inlinecite{zha07a} have analyzed 
realistic solar convection simulations and found satisfactory inversion 
results for the shallow layers covered by the simulations. Validations of the
wave-speed perturbation inversions based on numerical simulations with 
preset structures have also been performed. Meanwhile, 
numerical simulations for magnetic structures with flows are also under 
development \cite{rem09, ste11, kit11}. Validations of the time-distance 
helioseismology techniques will be carried out as well using these simulations.

Cross-comparisons between different helioseismology 
techniques, \eg\ comparing the mean rotation speed from our pipeline analysis 
with global helioseismology results, and comparing the subsurface flow fields 
with results from ring-diagram analyses, will also be important for 
the validation.

\begin{table}
\caption{Measured mean travel times and uncertainties for both quiet 
Sun regions and active regions.}
\label{time_error}
\begin{tabular}{cccc}
\hline
annulus No. & mean travel time & uncertainty for quiet regions & uncertainty
for active regions \\
            & (min) & (min) & (min) \\
\hline
1 & 11.87 & 0.062 & 0.17 \\
2 & 18.82 & 0.061 & 0.25 \\
3 & 21.76 & 0.11 & 0.26 \\
4 & 25.85 & 0.11 & 0.19 \\
5 & 28.69 & 0.14 & 0.15 \\
6 & 31.18 & 0.14 & 0.14 \\
7 & 35.07 & 0.10 & 0.10 \\
8 & 38.86 & 0.12 & 0.11 \\
9 & 42.46 & 0.11 & 0.093 \\
10 & 46.63 & 0.14 & 0.11 \\
11 & 50.26 & 0.15 & 0.11 \\
\hline
\end{tabular}
\end{table}

\subsection{Error Estimate} 
 \label{S33}
There are two types of errors in the pipeline results: systematic errors
due to our limited knowledge of the wave physics in the magnetized turbulent
medium and simplified mathematical formulations, and statistical errors,
which are mostly due to the stochastic nature of the solar oscillations
(``realization noise"). Here, we only discuss the statistical errors.

To estimate the errors in the inversion results, we need first to estimate
the uncertainties in the measured acoustic travel times. Here, we estimate
the measurement uncertainties following the prescriptions of 
\inlinecite{jen03} and \inlinecite{cou06}. We select 25 quiet-Sun regions, 
and use the rms variation of mean travel times for different measurement 
distances as an error estimate for the travel times. The estimated uncertainties for the Gabor wavelet fitting are given in Table~\ref{time_error}, and the 
uncertainties obtained for the GB algorithm are similar, but slightly 
larger for short distances and slightly smaller for long distances.
Active regions have different measurement uncertainties. To estimate these,
we selected a relatively stable sunspot, NOAA AR~11092, from 2 through 5 
August 2010, and we assumed that the sunspot did not change during this
period. Then we use the rms of the travel times measured inside the sunspot 
as an error estimate. Due to the evolution of the sunspot, this approach 
overestimates the measurement uncertainties, but can still give us an
approximate estimate of measurement errors. These error estimates are 
presented in Table~\ref{time_error} as well.

Inversions are then performed for the same quiet-Sun regions and the active 
region to estimate the statistical errors in inversion results. 
Following the same approach, the rms of inverted wave-speed 
perturbations is assumed as the statistical error. The error 
estimates for both the quiet Sun and active regions are
shown in Table~\ref{inver_err}. Because supergranular flows are 
dominant in the flow fields, it is difficult to estimate errors of the inverted
velocity for the quiet Sun by this approach. Instead, we estimate the 
velocity errors based on the rotational velocity profile, which has 
little change in a time scale of a few days. 
While the errors for the wave-speed perturbation
inside active regions are roughly twice of those for the quiet-Sun regions, 
the velocity errors inside active regions sometimes are 
seven times larger than the errors in the quiet Sun.

\begin{table}
\caption{Error estimates for the relative wave-speed perturbation and 
horizontal velocity inferences in the quiet-Sun (QS) and active regions (AR).}
\label{inver_err}
\begin{tabular}{ccccc}
\hline
Depth & wave speed for QS & velocity for QS & wave speed for AR & 
velocity for AR \\
(Mm)  &  & (m s$^{-1}$) & & (m s$^{-1}$) \\
\hline
0 -- 1 & $2.8\times10^{-3}$ & 7.8 & $6.3\times10^{-3}$ & 58.3 \\
1 -- 3 & $4.1\times10^{-3}$ & 7.5 & $10.9\times10^{-3}$ & 56.4 \\
3 -- 5 & $6.4\times10^{-3}$ & 8.9 & $8.7\times10^{-3}$ & 51.1 \\
5 -- 7 & $4.6\times10^{-3}$ & 9.4 & $9.7\times10^{-3}$ & 45.1 \\
7 -- 10 & $4.7\times10^{-3}$ & 13.1 & $6.7\times10^{-3}$ & 34.5 \\
10 -- 13 & $3.7\times10^{-3}$ & 12.9 & $3.1\times10^{-3}$ & 28.1 \\
\hline
\end{tabular}
\end{table}

\section{Data Products and Initial Results from HMI}
 \label{S4}
The time--distance data analysis pipeline is used for the routine production of 
nearly real-time full-disk (actually, nearly full-disk covering a
$120\degr\times120\degr$ area on the solar disk) wave-speed perturbation 
and flow field maps every eight hours. These maps are then used to construct 
the corresponding synoptic maps for each Carrington rotation. The pipeline 
can also be used for specific target areas, such as active regions.
In this section, we introduce the data products from this pipeline and 
some initial results from it.

\begin{figure}
\centerline{\includegraphics[width=0.8\textwidth]{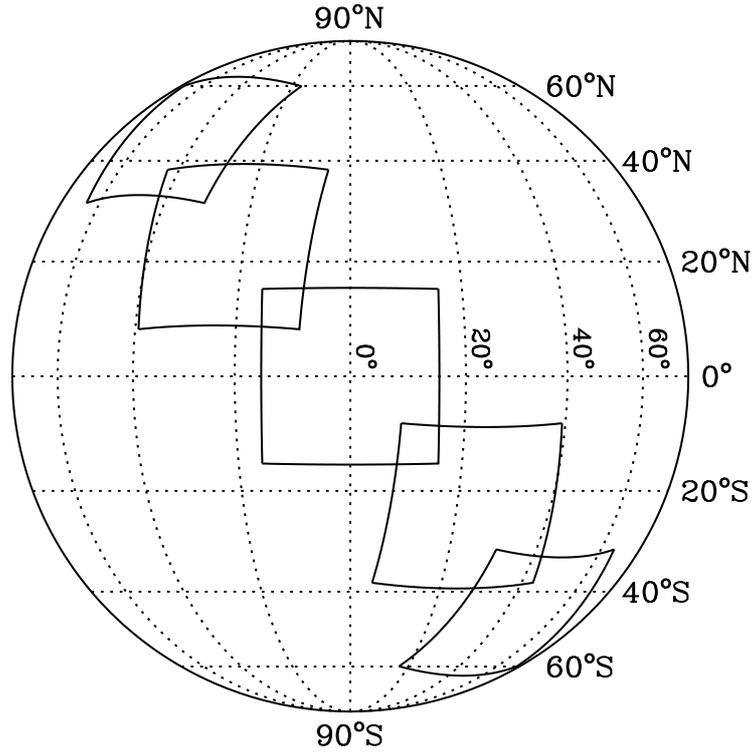} }
\caption{Schematic plot showing how areas are selected for a routine 
calculations of the full-disk wave-speed and flow maps. Not all of the 
25 selected areas are shown. }
\label{schematic}
\end{figure}

\subsection{Routine Production: Full-Disk and Synoptic Maps}
 \label{S41} 
For each day of HMI observations, we select three eight-hour periods: 00:00 
-- 07:59 UT, 08:00 -- 15:59 UT, and 16:00 -- 23:59 UT. For each analysis period,
we select 25 regions, with the central locations at $0\degr$, $\pm24\degr$,
and $\pm48\degr$ in both longitude and latitude, where the longitude
is relative to the central meridian at the mid-time of the selected period. 
Figure~\ref{schematic} shows locations of these areas on the solar disk. 
The total number of areas is 25: five rows and five columns. 
Due to the Postel's projection, the boundaries of these areas are often 
not parallel to the latitude or longitude lines. It is also evident that 
many areas overlap, some areas overlap twice, and
some overlap four times. The travel times and inversion results 
are averaged in these overlapped areas. However, in the areas close 
to the solar limb, the foreshortening effect may become non-negligible, 
but the role of this is not yet systematically studied. Users 
of these maps are urged to be cautious when using the pipeline
results in the areas close to the limb.

\begin{figure}
\centerline{\includegraphics[width=0.90\textwidth]{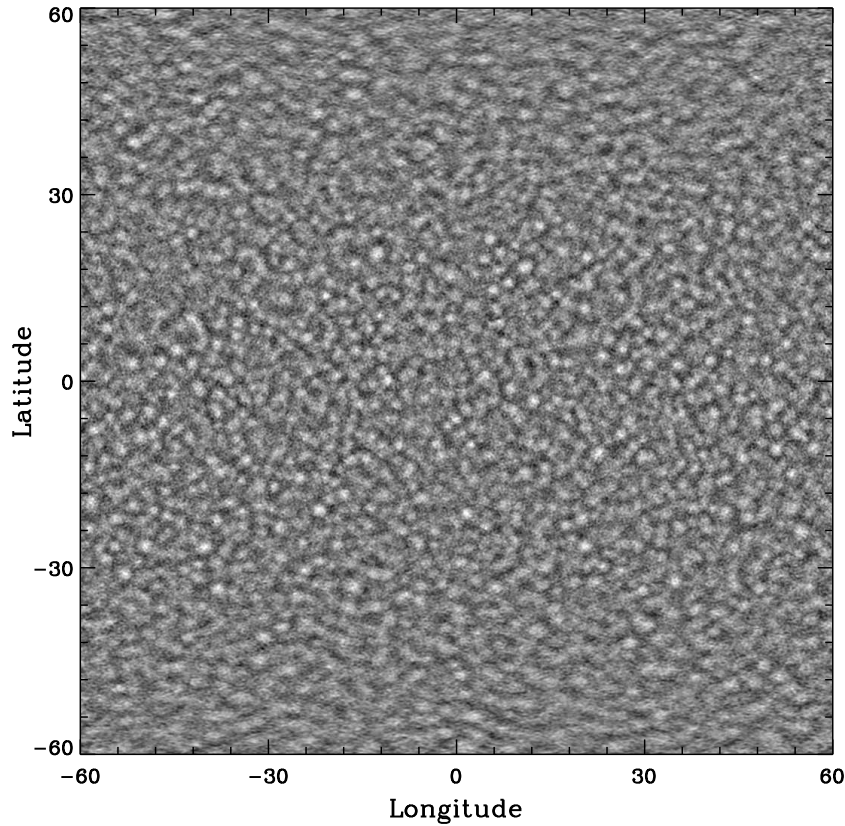} }
\caption{A map of horizontal flow divergence for a depth range of 1 -- 3 Mm 
and a time period of 00:00 -- 07:59 UT 19 May 2010. The display scale is from 
$-6.2\times10^{-4}$ to $6.2\times10^{-4}$ s$^{-1}$. White areas with positive
divergence correspond to supergranulation. Note that supergranules appear 
larger at high latitudes because of the rectangular longitude--latitude 
map projection.}
\label{fd_div}
\end{figure}

\begin{figure}
\centerline{\includegraphics[width=0.80\textwidth]{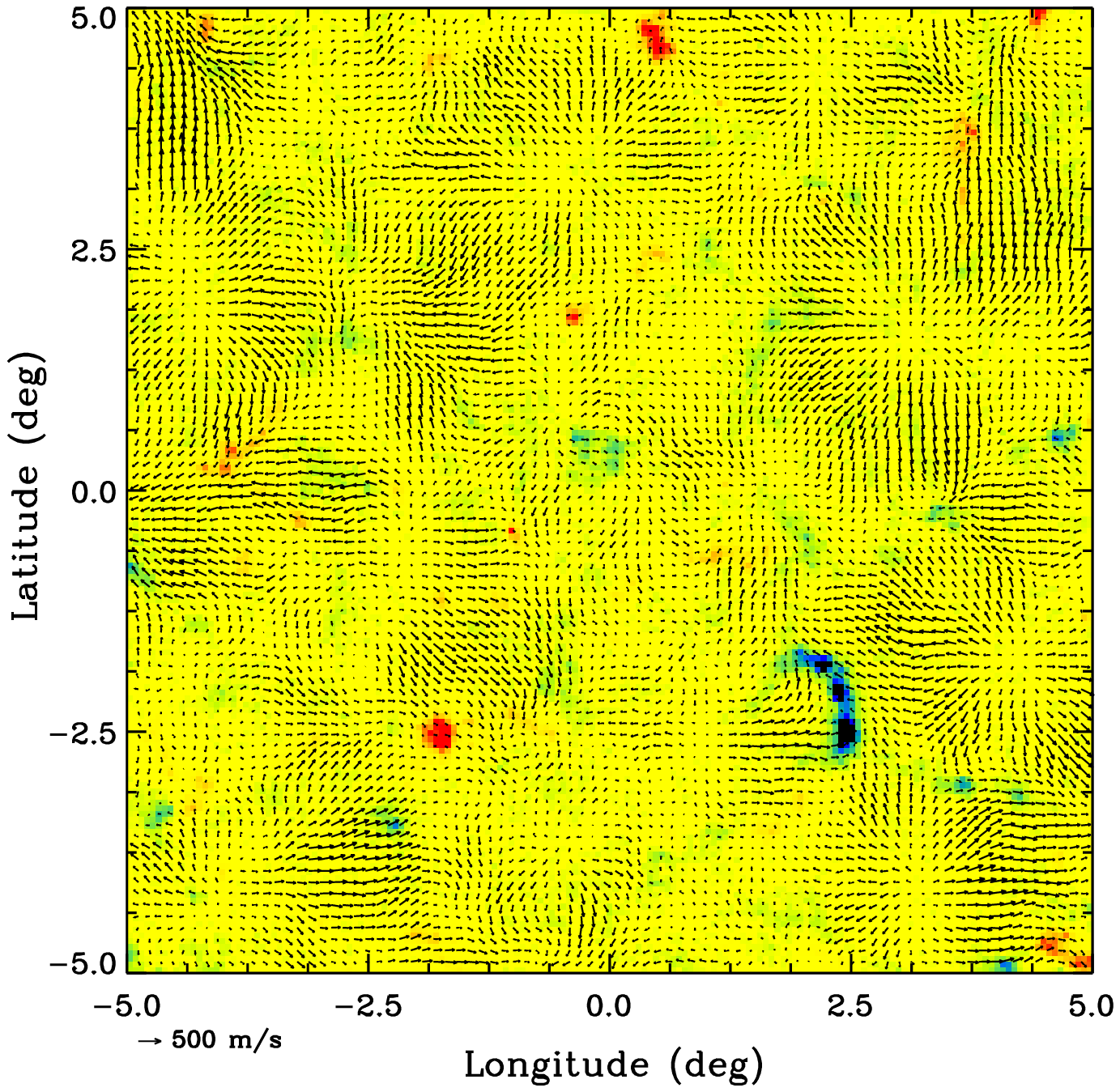} }
\caption{A sample of subsurface horizontal flow fields with full spatial 
resolution at the depth of 0 -- 1 Mm. This area is sampled at the center of 
the map shown in Figure~\ref{fd_div}. The background image shows the 
line-of-sight magnetic field measured by HMI, with red as positive and 
blue as negative polarities. The range of the magnetic field is from 
$-80$ to 80 Gauss. }
\label{super_flow}
\end{figure}

For each full-disk map and each synoptic map, the East -- West 
velocity $[v_x]$, the North -- South velocity $[v_y]$, and wave-speed 
perturbation $[\delta c/c]$ in each depth layer are derived with 
a horizontal spatial sampling of $0.12\degr$ pixel$^{-1}$. For each
of the 25 areas, the inversion results are first obtained in the 
Postel-projection coordinates, and then converted into the longitude--latitude
coordinates. This coordinate conversion is basically the inverse
procedure of transforming the observed data into the Postel-projection
coordinates for the travel-time measurements. Cubic spline interpolation 
is employed. The results in high latitude regions are oversampled.
After the coordinate transformation, the overlapped areas are averaged,
and the final statistical errors are estimated for the whole procedure starting
from the travel-time measurements. This includes all potential errors from
the interpolation and remapping. The final full-disk results are saved 
on a uniform longitude--latitude coordinate grid, so each horizontal
image of the subsurface layers has a total of $1000\times1000$ pixels 
covering $120\degr$ 
in both longitudinal and latitudinal directions. This coordinate choice
is convenient for merging and analyzing results, but unavoidably 
distorts maps in high latitude areas. 

\begin{figure}
\includegraphics[width=0.83\textwidth]{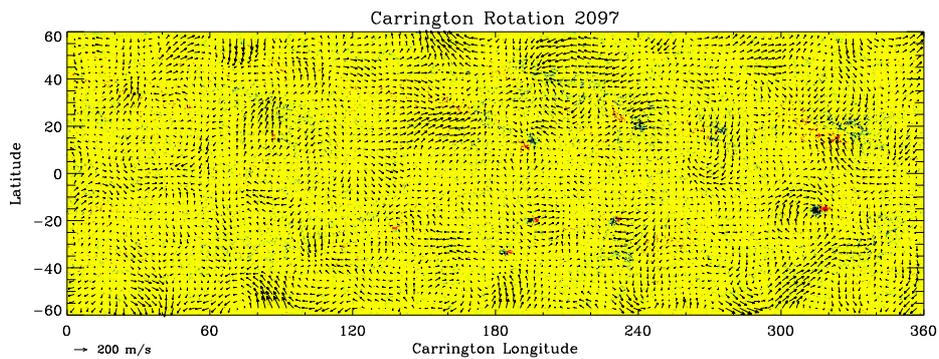} 
\caption{Synoptic map of large-scale horizontal flows at the depth of 
$1 - 3$ Mm for CR 2097. The background image is the corresponding 
line-of-sight magnetic field, with red as positive and blue as negative 
polarities, in the range of $-50$ to 50 Gauss. }
\label{synop}
\end{figure}

Figure~\ref{fd_div} shows an example of a full-disk map of the subsurface 
horizontal flow divergence calculated from $v_x$ and $v_y$ at the depth range 
of 1 -- 3 Mm. The positive divergence areas correspond to supergranulation. 
Such maps at various depths with continuous temporal coverage can be 
valuable for studying the structure and evolution of the supergranulation.
Figure~\ref{super_flow} displays the subsurface horizontal flow fields
with full spatial resolution for a region located at the center of the map 
in Figure~\ref{fd_div}. Supergranular flows can be easily identified. 

\begin{figure}
\centerline{\includegraphics[width=0.70\textwidth]{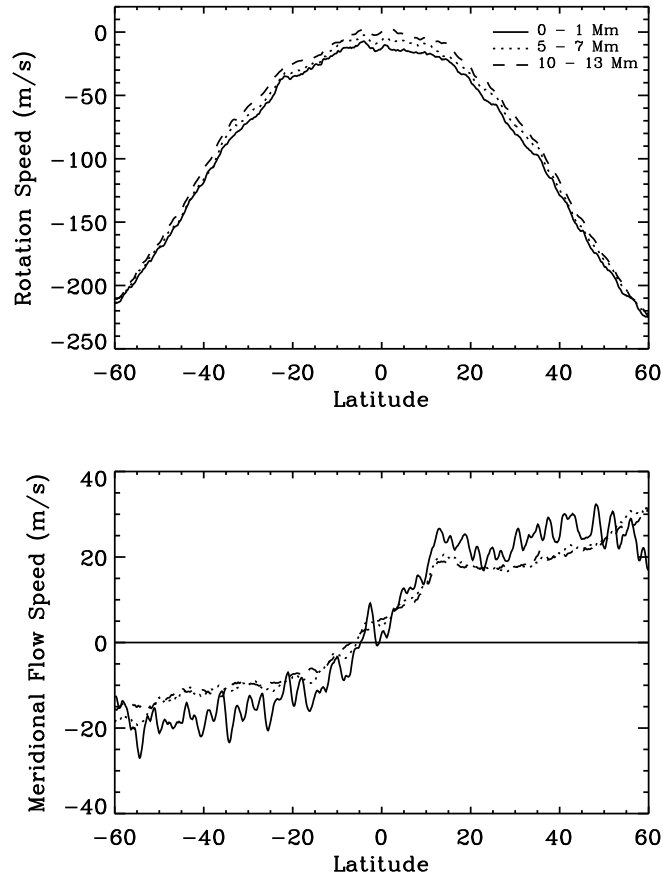} }
\caption{Averaged rotation (upper) and meridional flow speeds (lower) 
at selected depths for Carrington Rotation 2097. The rotation 
speed is relative to the constant Carrington rotation rate. }
\label{rot_mer}
\end{figure}

From the full-disk wave-speed perturbation and flow maps obtained every eight 
hours, we select an area $13.2\degr$ wide in longitude, {\it i.e.} 
$-6.6\degr$ to $6.6\degr$ from the central meridian, to construct the synoptic 
wave-speed perturbation and flow maps. Since the Carrington rotation 
rate corresponds to a shift of approximately $4.4\degr$ every eight hours, 
each location in the constructed synoptic maps is averaged roughly three 
times ({\it i.e.} one whole day). The resultant synoptic 
map for each depth consists of $3000\times1000$ pixels. Since such a map is
difficult to display, we show in Figure~\ref{synop} a binned-down
synoptic large-scale flow map obtained for the depth of $1 - 3$ Mm
for Carrington Rotation 2097 during the period from 20 May to 16 June, 
2010. The vectors in the figure are obtained 
by averaging the flow fields in areas of $15\degr\times15\degr$ with a 
$3\degr$ sampling rate. The maps of this type are similar to the 
subsurface flow maps obtained from the ring-diagram analysis \cite{hab02}.
From Figure~\ref{synop}, it can be found that the large-scale flows 
often converge around magnetic regions. 

Normally, the full-disk flow maps and the synoptic flow maps are provided as
residual flow maps after subtracting the flows averaged over the entire
Carrington rotation. The subtracted average maps contain the differential 
rotation, meridional flows, and possibly some systematic effects.
Figure~\ref{rot_mer} shows 
examples of the subsurface differential rotation speed and meridional 
flow speed obtained by averaging the calculations for Carrington Rotation 
2097. These results are also provided online together with the synoptic 
flow maps. 

\begin{figure}
\centerline{\includegraphics[width=0.8\textwidth]{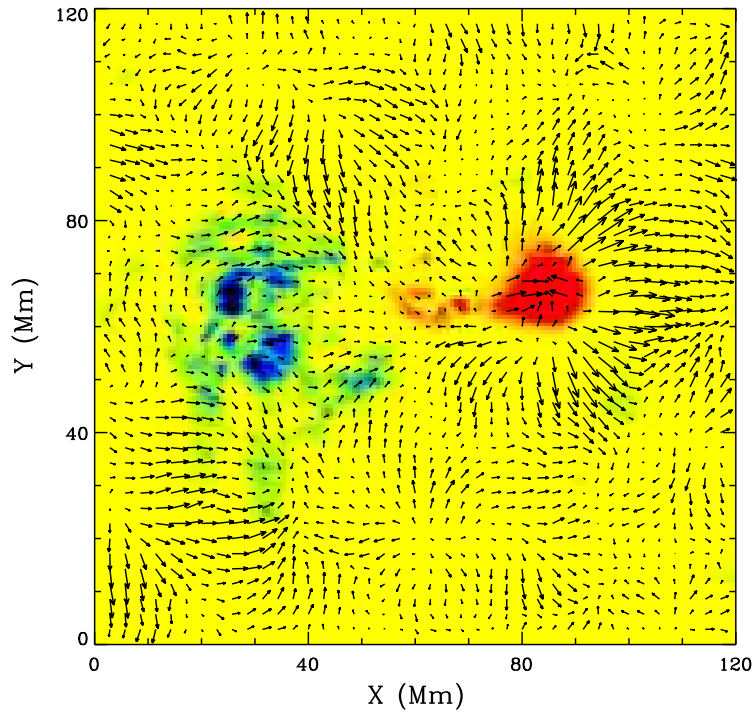} }
\caption{Subsurface flow fields of AR 11072 at the depth of $1 - 3$ Mm,
obtained from 16:00 -- 23:59 UT 23 May 2010. 
The background image is line-of-sight magnetic field, with red positive and 
blue negative. The image is displayed with a scale from $-1000$ to $1000$ 
Gs. The arrows are displayed after a $2\times2$ rebinning, and the longest 
arrow represents a speed of 300 m~s$^{-1}$. }
\label{spot}
\end{figure}

\subsection{Target Areas}
 \label{S42}
As already mentioned in Section~\ref{S2}, the pipeline can also be run for 
specific target areas and specific time intervals. The pipeline users are 
required to
provide the Carrington coordinates of the center of the target area, and 
the mid-time of the time interval. 

Figure~\ref{spot} shows an example of a small part (approximately 1/9) of 
a target area, with an active region, AR~11072, included in it. 
The subsurface flow 
field, at the depth of $1 - 3$ Mm, is displayed after a $2\times2$ rebinning. 
The displayed time period, 16:00 -- 23:59 UT on 23 May 2010, is approximately 
2.5 days after the start of emergence of this active region that
was still growing. Our results clearly
show subsurface outflows around the leading sunspot, and some converging
flows inside it. Comparing with the previous results of
\inlinecite{kos09} and \inlinecite{zha10}, one may conclude 
that the subsurface flow fields of active regions evolve with the 
evolution of active regions. There may be prominent outflows around 
sunspots during their early growing phase and also their decaying phase, but 
there may also be strong converging flows during the stable period. 
Our continuous 
monitoring of the solar full-disk subsurface flows may give us an 
opportunity to statistically study the changes of sunspot's subsurface 
flows with the sunspot evolution.

\section{Summary}
 \label{S5}
We have developed a time--distance helioseismology data analysis pipeline for SDO/HMI
Doppler observations. This pipeline performs acoustic travel-time 
measurements based on two different methods, and conducts inversions
based on two different sensitivity kernels calculated in the ray-path and 
Born approximations. The pipeline gives nearly real-time routine products 
of full-disk wave-speed perturbations and flow field maps in the range of
depth 0 -- 20 Mm every eight hours, and provides the corresponding synoptic 
wave-speed perturbation and flow field maps for each Carrington rotation. 
The averaged rotation speed and meridional flow speed are also provided 
separately for each rotation. In addition to these routine production, 
the pipeline can also be used for analysis of specific target areas 
for specific time intervals. This data analysis pipeline will provide 
important information about the subsurface structure and dynamics 
on both local and global scales, and its continuous coverage through 
years to come will be useful for understanding the connections 
between solar subsurface properties and magnetic activity in the chromosphere
and corona. With the improvement of our understanding of acoustic 
wave and magnetic field interactions, and also the measurement and 
inversion techniques, the pipeline codes will be regularly revised.

\end{article}

\begin{thebibliography}{}

\bibitem[\protect\citeauthoryear{{Beck}, {Gizon}, and {Duvall}}{2002}]{bec02}
        Beck, J.G., Gizon, L., Duvall, T.L. Jr.: 2002, \apjl{} \textbf{575}, 
        L47

\bibitem[\protect\citeauthoryear{{Birch} and {Gizon}}{2007}]{bir07}
        Birch, A.C., Gizon, L.: 2007, \an{} \textbf{328}, 228

\bibitem[\protect\citeauthoryear{{Birch} and {Kosovichev}}{2000}]{bir00}
        Birch, A.C., Kosovichev, A.G.: 2000, \solphys{} \textbf{192}, 193

\bibitem[\protect\citeauthoryear{{Birch}, {Kosovichev}, and {Duvall}}{2004}]
        {bir04} Birch, A.C., Kosovichev, A.G., Duvall T.L. Jr.: 2004,
        \apj{} \textbf{608}, 580

\bibitem[\protect\citeauthoryear{{Chou} and {Dai}}{2001}]{cho01} Chou, D.-Y.,
        Dai, D.-C.: 2001, \apjl{} \textbf{559}, L175

\bibitem[\protect\citeauthoryear{Couvidat \etal}{2006}]{cou06} Couvidat, S.,
        Birch, A.C., Kosovichev, A.G.: 2006, \apj{} \textbf{640}, 516

\bibitem[\protect\citeauthoryear{Couvidat \etal}{2004}]{cou04} Couvidat, S.,
        Birch, A.C., Kosovichev, A.G., Zhao, J.: 2004, \apj{} \textbf{607}, 554

\bibitem[\protect\citeauthoryear{Couvidat \etal}{2005}]{cou05} Couvidat, S.,
        Gizon, L., Birch A.C., Larsen, R.M., Kosovichev, A.G.: 2005,
        {\it Astrophy. J. Supp. Ser.} \textbf{158}, 217

\bibitem[\protect\citeauthoryear{Couvidat \etal}{2011}]{cou11} Couvidat, S.,
        Zhao, J., Birch, A.C., Kosovichev, A.G., Duvall, T.L. Jr., 
        Parchevsky, K.V., Scherrer, P.H.: 2010, \solphys{} 
        \url{DOI: 10.1007/s11207-010-9652-y}

\bibitem[\protect\citeauthoryear{Duvall \etal}{1996}]{duv96} Duvall, T.L. Jr.,
        D'Silva, S., Jefferies, S.M., Harvey, J.W., Schou, J.: 1996,
        \nat{} \textbf{379}, 235

\bibitem[\protect\citeauthoryear{{Duvall} and {Gizon}}{2000}]{duv00} 
        Duvall T.L. Jr., Gizon, L.: 2000, \solphys{} \textbf{192}, 177

\bibitem[\protect\citeauthoryear{Duvall \etal}{1993}]{duv93} Duvall, T.L. Jr.,
        Jefferies, S.M., Harvey, J.W., Pomerantz, M.A.: 1993, \nat{}
        \textbf{362}, 430

\bibitem[\protect\citeauthoryear{Duvall \etal}{1997}]{duv97} Duvall, T.L. Jr.,
        Kosovichev, A.G., Scherrer, P.H., Bogart, R.S., Bush, R.I., DeForest,
        C., Hoeksema, J.T., Schou, J., Saba, J.L.R., Tarbell, T.D., Title, A.M.,
        Wolfson, C.J., Milford, P.N.: 1997, \solphys{} \textbf{170}, 63

\bibitem[\protect\citeauthoryear{Georgobiani \etal}{2007}]{geo07} 
        Georgobiani, D., Zhao, J., Kosovichev, A.G., Bensen, D., Stein, R.F., 
        Nordlund, \AA.: 2007, \apj{} \textbf{657}, 1157

\bibitem[\protect\citeauthoryear{Giles \etal}{1997}]{gil97} Giles, P.M.,
        Duvall, T.L. Jr., Scherrer, P.H., Bogart, R.S.: 1997, \nat{}
        \textbf{390}, 52

\bibitem[\protect\citeauthoryear{{Gizon} and {Birch}}{2002}]{giz02} Gizon, L.,
        Birch, A.C.: 2002, \apj{} \textbf{571}, 966


\bibitem[\protect\citeauthoryear{{Gizon}, {Duvall}, and {Larsen}}{2000}]
        {giz00} Gizon, L., Duvall, T.L. Jr., Larsen, R.M.: 2000, 
        {\it J. Astrophys. Astron.} \textbf{21}, 339

\bibitem[\protect\citeauthoryear{Gizon \etal}{2009}]{giz09} Gizon, L., 
        Schunker, H., Baldner, C.S., Basu, S., Birch, A.C., Bogart, R.S.,
        Braun, D.C., Cameron, R., Duvall, T.L. Jr., Hanasoge, S.M., 
        Jackiewicz, J., Roth, M., Stahn, T., Thompson, M.J., Zharkov, S.:
        2009, \ssr{} \textbf{144}, 249

\bibitem[\protect\citeauthoryear{Haber \etal}{2002}]{hab02} Haber, D.A., 
        Hindman, B.W., Toomre, J., Bogart, R.S., Larsen, R.M., Hill, F.:
        2002, \apj{} \textbf{570}, 885

\bibitem[\protect\citeauthoryear{Hanasoge \etal}{2008}]{han08} Hanasoge, S.M.,
        Couvidat, S., Rajaguru, S.P., Birch, A.C.: 2008, \mnras{} \textbf{391},
        1931

\bibitem[\protect\citeauthoryear{{Ilonidis}, {Zhao}, and {Hartlep}}{2009}]
        {ilo09} Ilonidis, S., Zhao, J., Hartlep, T.: 2009, \solphys{}
        \textbf{258}, 181

\bibitem[\protect\citeauthoryear{Jacobsen \etal}{1999}]{jac99} Jacobsen, B.H., 
        M{\o}ller, I., Jensen, J.M., Effers{\o}, F.: 1999, {\it Phys. Chem. 
        of the Earth A - Solid Earth and Geodesy} \textbf{24}, 15

\bibitem[\protect\citeauthoryear{{Jackiewicz}, {Gizon}, and {Birch}}{2008}]
        {jac08} Jackiewicz, J., Gizon, L., Birch, A.C.: 2008, \solphys{}
        \textbf{251}, 381

\bibitem[\protect\citeauthoryear{Jensen \etal}{2003}]{jen03} Jensen, J.M.,
        Duvall, T.L. Jr., Jacobsen, B.H.: 2003, in Proc. {\it SOHO} 12/
        GONG+ 2002, Local and Global Helioseismology: the Present and 
        Future, ed. H. Sawaya-Lacoste (ESA SP-517; Noordwijk), 315

\bibitem[\protect\citeauthoryear{Jensen \etal}{2001}]{jen01} Jensen, J.M.,
        Duvall, T.L. Jr., Jacobsen, B.H., Christensen-Dalsgaard, J.: 2001,
        \apjl{} \textbf{553}, L193

\bibitem[\protect\citeauthoryear{{Jensen}, {Jacobsen}, and 
        {Christensen-Dalsgaard}}{2000}]{jen00} Jensen, J.M., Jacobsen, B.H.,
        Christensen-Dalsgaard, J.: 2000, \solphys{} \textbf{192}, 231

\bibitem[\protect\citeauthoryear{Kitiashvili \etal}{2011}]{kit11} Kitiashvili,
        I.N., Kosovichev, A.G., Mansour, N.N., Wray, A.A.: 2011, \solphys{}
        \textbf{268}, 283

\bibitem[\protect\citeauthoryear{Kosovichev}{1996}]{kos96} Kosovichev, A.G.:
        1996, \apjl{} \textbf{461}, L55

\bibitem[\protect\citeauthoryear{Kosovichev}{2009}]{kos09} Kosovichev, A.G.:
        2009, \ssr{} \textbf{144}, 175

\bibitem[\protect\citeauthoryear{Kosovichev}{2010}]{kos10} Kosovichev, A.G.:
        2010, arXiv: 1010.4927, \url{http://arxiv.org/abs/1010.4927}

\bibitem[\protect\citeauthoryear{{Kosovichev} and {Duvall}}{1997}]{kos97}
        Kosovichev, A.G., Duvall, T.L. Jr.: 1997, in: Pijers, F.P.,
        Christensen-Dalsgaard, J., Rosenthal, C.S. (eds.) {\it Proceedings
        of SCORe'96: Solar Convection and Oscillations and Their 
        Relationship}, Kluwer, Astrophysics and Astronomy Library \textbf{225},
        Dordrecht, 241

\bibitem[\protect\citeauthoryear{{Kosovichev}, {Duvall}, and {Scherrer}}
        {2000}]{kos00} Kosovichev, A.G., Duvall, T.L. Jr., Scherrer, P.H.:
        2000, \solphys{} \textbf{192}, 159

\bibitem[\protect\citeauthoryear{{Lindsey} and {Braun}}{2005a}]{lin05a}
        Lindsey, C., Braun, D.C.: 2005a, \apj{} \textbf{620}, 1107

\bibitem[\protect\citeauthoryear{{Lindsey} and {Braun}}{2005b}]{lin05b}
        Lindsey, C., Braun, D.C.: 2005b, \apj{} \textbf{620}, 1118

\bibitem[\protect\citeauthoryear{{Nigam} and {Kosovichev}}{2010}]{nig10}
        Nigam, R., Kosovichev, A.G.: 2010, \apj{} \textbf{708}, 1475


\bibitem[\protect\citeauthoryear{{Parchevsky}, {Zhao}, and {Kosovichev}}
        {2008}]{par08} Parchevsky, K.V., Zhao, J., Kosovichev, A.G.: 2008,
        \apj{} \textbf{678}, 1498

\bibitem[\protect\citeauthoryear{Rajaguru \etal}{2006}]{raj06} Rajaguru, S.P.,
        Birch, A.C., Duvall, T.L. Jr., Thompson, M.J., Zhao, J.: 2006,
        \apj{} \textbf{646}, 543

\bibitem[\protect\citeauthoryear{{Rempel}, {Sch\"{u}ssler}, and {Kn\"{o}lker}}
        {2009}]{rem09} Rempel, M., Sch\"{u}ssler, M., Kn\"{o}lker, M.: 2009,
        \apj{} \textbf{691}, 640

\bibitem[\protect\citeauthoryear{Sekii \etal}{2007}]{sek07} Sekii, T., 
        Kosovichev, A.G., Zhao, J., Tsuneta, S., Shibahashi, H., Berger, T.E.,
        Ichimoto, K., Katsukawa, Y., Lites, B., Nagata, S., Shimizu, T., 
        Shine, R.A., Suematsu, Y., Tarbell, T.D., Title, A.M.:
        2007, \pasj{} \textbf{59}, S637

\bibitem[\protect\citeauthoryear{Scherrer \etal}{1995}]{sch95} Scherrer, P.H.,
        Bogart, R.S., Bush, R.I., Hoeksema, J.T., Kosovichev, A.G., Schou, J.,
        Rosenberg, W., Springer, L., Tarbell, T.D., Title, A., Wolfson, C.J.,
        Zayer, I., MDI Engineering Team: 1995, \solphys{} \textbf{162}, 129

\bibitem[\protect\citeauthoryear{Schou \etal}{2011}]{sch11} Schou, J., 
        Scherrer, P.H., Watchter, R., Couvidat, S., Bush, R.I., \etal:
        2011, \solphys{}, in preparation 

\bibitem[\protect\citeauthoryear{Schunker \etal}{2005}]{sch05} Schunker, H.,
        Braun, D.C., Cally, P.S., Lindsey, C.: 2005, \apjl{} \textbf{621}, L149

\bibitem[\protect\citeauthoryear{{Snodgrass} and {Ulrich}}{1990}]{sno90}
        Snodgrass, H.B., Ulrich, R.K.: 1990, \apj{} \textbf{351}, 309

\bibitem[\protect\citeauthoryear{{Stein \etal}}{2011}]{ste11}
        Stein, R.F., Lagerfj\"{a}rd, A., Nordlund, \AA., Georgobiani, D.:
        2011, \solphys{} \textbf{268}, 271

\bibitem[\protect\citeauthoryear{Zhao}{2007}]{zha07b} Zhao, J.: 2007, \apjl{}
        \textbf{664}, L139

\bibitem[\protect\citeauthoryear{Zhao \etal}{2007}]{zha07a} Zhao, J., 
        Georgobiani, D., Kosovichev, A.G., Benson, D., Stein, R.F.,
        Nordlund, \AA.: 2007, \apj{} \textbf{659}, 848

\bibitem[\protect\citeauthoryear{Zhao \etal}{2009}]{zha09} Zhao, J., 
        Hartlep, T., Kosovichev, A.G., Mansour, N.N.: 2009, \apj{} 
        \textbf{702}, 1150


\bibitem[\protect\citeauthoryear{{Zhao} and {Kosovichev}}{2004}]{zha04}
        Zhao, J., Kosovichev, A.G.: 2004, \apj{} \textbf{603}, 776

\bibitem[\protect\citeauthoryear{{Zhao} and {Kosovichev}}{2006}]{zha06}
        Zhao, J., Kosovichev, A.G.: 2006, \apj{} \textbf{643}, 1317

\bibitem[\protect\citeauthoryear{{Zhao}, {Kosovichev}, and {Duvall}}{2001}]
        {zha01} Zhao, J., Kosovichev, A.G., Duvall, T.L. Jr.: 2001, \apj{}
        \textbf{557}, 384

\bibitem[\protect\citeauthoryear{{Zhao}, {Kosovichev}, and {Sekii}}{2010}]
        {zha10} Zhao, J., Kosovichev, A.G., Sekii, T.: 2010, \apj{}
        \textbf{708}, 304


\bibitem[\protect\citeauthoryear{{Zhao} and {Jordan}}{1998}]{zha98}
        Zhao, L., Jordan, T.H.: 1998, {\it Geophys. J. Int.} \textbf{133}, 683

\bibitem[\protect\citeauthoryear{{Zharkov} and {Thompson}}{2008}]{zha08}
        Zharkov, S., Thompson, M.J.: 2008, \solphys{} \textbf{251}, 369

\end{thebibliography}
\end{document}